\documentclass[aps,pre,superscriptaddress,floatfix,twocolumn]{revtex4-2}
\usepackage{amsfonts,color}
\usepackage{amsmath}
\usepackage{amssymb}
\usepackage{graphicx,dsfont}
\usepackage{physics}
\usepackage[utf8]{inputenc}
\usepackage{ulem}
\newlength{\ldag}
\newcommand{\element}[3]{\langle #1|#2|#3\rangle}
\newcommand{\Spin}{\hat{\mathcal{S}}}
\begin{document}

\title{Excited-State Quantum Phase Transitions in the anharmonic Lipkin-Meshkov-Glick model: Dynamical aspects}

\author{J. Khalouf-Rivera}
\affiliation{Departamento de F\'isica Aplicada III, Escuela T\'ecnica Superior de Ingenier\'ia, Universidad de Sevilla, 41092 Sevilla, Spain }
\affiliation{Departamento de Ciencias Integradas y Centro de Estudios Avanzados en Física, Matemáticas y Computación, Universidad de Huelva, Huelva 21071, Spain}

\author{J. Gamito}
\affiliation{Departamento de F\'isica At\'omica, Molecular y Nuclear, Facultad de F\'isica, Universidad de Sevilla, Apartado 1065, E-41080 Sevilla, Spain}

\author{F. Pérez-Bernal}
\affiliation{Departamento de Ciencias Integradas y Centro de Estudios Avanzados en Física, Matemáticas y Computación, Universidad de Huelva, Huelva 21071, Spain}
\affiliation{Instituto Carlos I de F\'{\i}sica Te\'orica y Computacional, Universidad de Granada, Fuentenueva s/n, 18071 Granada, Spain \\}

\author{J.M. Arias}
\affiliation{Departamento de F\'isica At\'omica, Molecular y Nuclear, Facultad de F\'isica, Universidad de Sevilla, Apartado 1065, E-41080 Sevilla, Spain}
\affiliation{Instituto Carlos I de F\'{\i}sica Te\'orica y Computacional, Universidad de Granada, Fuentenueva s/n, 18071 Granada, Spain \\}

\author{P. P\'erez-Fern\'andez}
\affiliation{Departamento de F\'isica Aplicada III, Escuela T\'ecnica Superior de Ingenier\'ia, Universidad de Sevilla, 41092 Sevilla, Spain }
\affiliation{Instituto Carlos I de F\'{\i}sica Te\'orica y Computacional, Universidad de Granada, Fuentenueva s/n, 18071 Granada, Spain \\}

\begin{abstract}
  The standard Lipkin-Meshkov-Glick (LMG) model undergoes a
  second-order ground-state quantum phase transition (QPT) and an
  excited-state quantum phase transition (ESQPT). The inclusion of an
  anharmonic term in the LMG Hamiltonian gives rise to a second ESQPT
  that alters the static properties of the model 
  [\textit{Phys. Rev. E} \textbf{106}, 044125 (2022)]. In the present work,
  the dynamical implications associated to this new ESQPT are
  analyzed. For that purpose, a quantum quench protocol is defined on
  the system Hamiltonian that takes an initial state, usually the
  ground state, into a complex excited state that evolves on time. The
  impact of the new ESQPT on the time evolution of the survival
  probability and the local density of states after the quantum quench, as
  well as on the Loschmidt echoes and the microcanonical out-of-time-order correlator (OTOC)
  are discussed. The anharmonity-induced ESQPT, despite having a
  different physical origin, has dynamical consequences similar to
those observed in the ESQPT already present in the standard LMG model.
\end{abstract}

\maketitle

\section{Introduction}

The use of toy models has been fundamental for important
advances in all branches of Physics. These are nontrivial models but still simple enough to be
solved analytically and they can be used either to
look into limiting situations in complex systems or to check and
better understand different approximation techniques. Some relevant
examples of solvable models are Elliott's rotational su(3) model
\cite{Elliott} and the Interacting Boson Model \cite{Arima1975,Arima1976,Arima1978,Arima1979} in Nuclear
Physics, the Rabi \cite{RabiI,RabiII}, Jaynes-Cummings \cite{JC} and Dicke
models \cite{Dicke} in Quantum Optics, or the Lipkin-Meshkov-Glick
(LMG) model in many-body physics \cite{Lipkin1,Lipkin2,Lipkin3}, just
to mention few of them. In many cases, such models were originally introduced in a particular branch of Physics and
they were later used in completely different fields. In
particular, the LMG model was originally proposed to test
many-body approximations such as the time-dependent Hartree-Fock or
perturbation methods in nuclear systems
\cite{Lipkin1,Lipkin2,Lipkin3}, but it has demonstrated to be very useful for
the study of quantum phase transitions (QPTs)
\cite{Castanos2005,QPT1,QPT2,Romera2014} and has been realized
experimentally with optical cavities \cite{Morrison2008},
Bose-Einstein condensates \cite{Zibold2010}, nuclear magnetic
resonance systems \cite{AFerreira2013}, trapped atoms
\cite{Jurcevic2014, Jurcevic2017, Muniz2020, Li2022}, and cold atoms
\cite{Makhalov2019}. For instance, the LMG model has been used to test
the possible existence of excited state quantum phase transitions
(ESQPTs) \cite{ESQPT} and relations between ESQPTs and quantum
entanglement \cite{Enta1,Enta2}, or quantum decoherence
\cite{Pedro1}. The ESQPT concept was introduced in
\cite{Cejnar-Iachello} and an excellent review on this topic has been
recently published \cite{Cejnar-review}.

\medskip

It is worth noting that phase transitions are well defined for
macroscopic systems, however, the same ideas can be applied in mesoscopic systems where one can observe phase transition
precursors even for moderate system sizes \cite{IZ04}. When
dealing with mesoscopic systems, the study of their mean field, or
large-size limit, is a
valuable reference to connect the precursors with the nonanaliticities expected in a QPT. Toy models, such as the LMG model,
are simple enough to be solved for a large number of particles,
allowing for a clear connection with the aforementioned large-size limit.

\medskip

This work is part of a more complete study on the anharmonic LMG
(ALMG) model. The additional anharmonic term induces, in addition to
the already known ESQPT \cite{Vidal,Pedro1}, an anharmonicity-induced
ESQPT that needs to be well understood. In a previous publication
\cite{Gamito1}, the static aspects of both the ground state QPT and
the two ESQPT’s in the ALMG model were characterized. A mean field
analysis in the large-N limit was performed and different observables
were used to characterize the different quantum phase transitions
involved: the energy gap between adjacent levels, the ground state QPT
order parameter, the participation ratio, the quantum fidelity
susceptibility, and the level density. In this work, we concentrate on
the influence of the two ESQPTs on the dynamics of the ALMG
model. With this aim, a quantum quench protocol that consists of an
abrupt change in one of the control parameters in the ALMG Hamiltonian
is defined. Then, the local density of states (LDOS, also known as strength function) together with the evolution of the system after the quench are
studied using the time evolution of the survival probability, Loschmidt echoes, and an out-of-time-order correlator (OTOC). 

\medskip

The present paper is organized as follows. In Sec.~II, the ALMG model
is introduced, its algebraic structure reviewed, and the relevant
matrix elements for the calculations in the $u(1)$ basis are
explicitly given. Sec.~III is devoted to the analysis of a quantum
quench protocol. Particularly, the time evolution of the survival
probability when the system undergoes a quantum quench is discussed to
understand how this quantity is influenced by the presence of the
ESQPTs in the system. In Sec.~IV, the ESQPTs impact on the evolution
of an OTOC is explored. Finally, some conclusions are presented in
Sec.~V.

\section{The model}

The LMG model can be used to describe one-dimensional spin-$1/2$
lattices with infinite-range interactions \cite{Lipkin1, Lipkin2,
  Lipkin3}. For an array of $N$ sites, the Hamiltonian is written in
terms of collective spin operators $\Spin_{\beta}=\sum_{i=1}^{N} \hat
s_{i,\beta}$ with $\beta=x,y,z$ and where $\hat s_{i,\beta}$ is the
$\beta$ component of the spin operator for a particle in site
$i$. Therefore, the usual LMG Hamiltonian is written as,

\begin{equation}
\hat H=(1-\xi)\left(S + \Spin_{z} \right)+\frac{2\xi}{S}\left(S^2-\Spin_{x}^2\right),
\label{eq:LMG1}
\end{equation}
with $S=N/2$. The operator $\Spin_{x}$ can be written in terms of the
usual ladder operators $\Spin_{+}$ and $\Spin_{-}$, defined as
$\Spin_{\pm}=\Spin_{x}\pm\imath \Spin_{y}$, and $\xi \in [0,1]$ is a
control parameter that drives the system from one phase to the other
one.  Indeed, from an algebraic point of view, the Eq.~\eqref{eq:LMG1}
LMG Hamiltonian presents a $u(2)$ algebraic structure with two
limiting dynamical symmetries: $u(2)\supset u(1)$ and $u(2)\supset
so(2)$ \cite{Frankbook}. Each dynamical symmetry is associated with a
different phase of the physical system. For $\xi=0$ the system reduces
to the $u(1)$ dynamical symmetry and this phase is usually referred to
as the normal (or symmetric) phase, whereas for $\xi=1$ the $so(2)$
dynamical symmetry is realized and the corresponding phase is called the
deformed (or broken-symmetry) phase \cite{Frankbook}.

Inspired by the works in Refs.~\cite{Curro-Osiris, KRivera2019,KRivera2021}, we
have included in the Eq.~\eqref{eq:LMG1} Hamiltonian a second-order
Casimir operator of $u(2)$, $S_{z}^2$,

\begin{equation}
\begin{split}
\hat H = & (1-\xi)\left(S + \Spin_{z} \right)
%+\frac{\alpha}{2S}\left(S+\Spin_z\right)\left(S+\Spin_z+1\right)
+\frac{2\xi}{S}\left(S^2-\Spin_{x}^2\right) \\ &
+\frac{\alpha}{2S}\left(S+\Spin_z\right)\left(S+\Spin_z+1\right)~.
\end{split}
\label{eq:LMG2}
\end{equation}

Again, the Hamiltonian depends on the $\xi$ control parameter which
drives the system between phases. In addition, a new control
parameter, $\alpha$, is introduced. The purpose of this work is to
explore the influence of this new term and the corresponding control
parameter on the dynamics of the system. It is worth noticing that for
$\alpha =0$, the original Hamiltonian, Eq.~\eqref{eq:LMG1}, is
recovered, and for $\alpha$ different from zero, the $\xi=0$ limit is
transformed from a truncated one-dimensional harmonic oscillator to an
anharmonic oscillator. That is the reason why Hamiltonian \eqref{eq:LMG2} is referred to as the anharmonic LMG
model. Moreover, we observe that the $so(2)$ limit is not longer
recovered for $\xi=1$ unless $\alpha$ is zero.

The Hilbert space for this system has dimension $2^N$, but due to the
conservation of the total spin, $[\Spin^2,\hat H]=0$, we can focus on
the sector of maximum irrep of the system, so the total spin quantum
number $S=N/2$ through the work. This leads to a drastic reduction of
Hilbert space dimension that now becomes $N+1$. On the other hand, the
basis for the Hilbert space given by the subalgebra $u(1)$,
$|S,M_z\rangle$ with $M_z=-N/2,...,0,...,N/2$ (the projection of the
total spin $S$ on the $z$ direction), is used along this work. The
matrix elements of Hamiltonian \eqref{eq:LMG2} in the $u(1)$
basis are given by
\begin{widetext}
\begin{eqnarray}
\element{S,M_{z}'}{\Spin_{z}}{S,M_{z}}&=&M_{z}\delta_{M_{z}',M_{z}}, \nonumber \\
\element{S,M_{z}'}{\Spin_{z}^{2}}{S,M_{z}}&=&M_{z}^{2}\delta_{M_{z}',M_{z}}, \nonumber \\
\Spin_{x}^2 &=& \frac{1}{4} ~\left(\Spin_{+}^2+\Spin_{-}^2+ \Spin_{+}\Spin_{-} + \Spin_{-}\Spin_{+}\right),
\nonumber \\
\element{S,M_{z}'}{\Spin_{+}\Spin_{-}+\Spin_{-}\Spin_{+}}{S,M_{z}}&=&\left(N\left(\frac{N}{2}+1\right)-2M_{z}^{2}\right)\delta_{M_{z}',M_{z}}, \nonumber \\
\element{S,M_{z}'}{\Spin_{+}^2}{S,M_{z}}&=& \sqrt{\frac{N}{2}(\frac{N}{2}+1)-M_{z}(M_{z}+1)}\sqrt{\frac{N}{2}(\frac{N}{2}+1)-(M_{z}+1)(M_{z}+2)})\delta_{M_{z}',M_{z+2}}, \nonumber \\
\element{S,M_{z}'}{\Spin_{-}^2}{S,M_{z}}&=& \sqrt{\frac{N}{2}(\frac{N}{2}+1)-M_{z}(M_{z}-1)}\sqrt{\frac{N}{2}(\frac{N}{2}+1)-(M_{z}-1)(M_{z}-2)})\delta_{M_{z}',M_{z-2}}.
\end{eqnarray}
\end{widetext}

In addition, Hamiltonian \eqref{eq:LMG2} conserves parity
$(-1)^{S+M_z}$ and the operator matrix can be split into two blocks, the first one including even
parity states  and the second one with odd parity states, with dimensions  $N/2+1$ and
dimension $N/2$ for an even $N$ value, .

A complete mean-field analysis of the semiclassical limit for
Hamiltonian \eqref{eq:LMG2} has been carried out using spin
coherent states in Ref.~\cite{Gamito1}, revealing for $\alpha < 0$ a
second order ground state QPT as well as two critical lines
corresponding to two ESQPTs and both marked by a high density of
states. A recently published work by Nader and collaborators focus on a general LMG Hamiltonian that can be easily connected with our ALMG realization \cite{Nader2021}. 
%They focus on the dynamics around allowed and avoided crossings in a region of parameter space where there is an ESQPT marked by a diverging density of states plus a different one with a step s different to the region we are exploring 
One of these high density of states critical lines was already
known for the LMG model \cite{Vidal,Pedro1}. Here, we pay heed to the other one, that we call
anharmonicity-induced ESQPT critical line \cite{Gamito1}. Particularly, it is worth
exploring whether this critical line is of a similar nature as the
other one and to what extent it has an impact on the system
dynamics. For this purpose, the dynamics of the system is studied by
means of the survival probability once the system undergoes a quantum
quench and an out-of-time-order correlator (OTOC).

\section{Quench dynamics}

The evolution of the system described by Hamiltonian \eqref{eq:LMG2}
after a quantum quench should be sensitive to the presence of
ESQPTs \cite{Pedro1, Pedro2, Pedro3, Santos2015, Nader2021}. We explore the ESQPT influence on the system dynamics with a quantum quench protocol, starting from
an eigenstate of the Hamiltonian, typically the ground state, and
following the system evolution once a control parameter in $\hat H$ is
abruptly modified. The quenching brings the system to an
excited state that evolves with time. The analysis of the ensuing
system dynamics is a valuable tool to detect and explore ESQPTs
in physical systems \cite{Pedro1,Pedro2}. Let us just note that, from a mathematical point of view,  this
quenching analysis can be put in relation to the survival probability or a particular realization of the Loschmidt echo.

\medskip

\begin{figure*}
\begin{centering}
\includegraphics[scale=0.70]{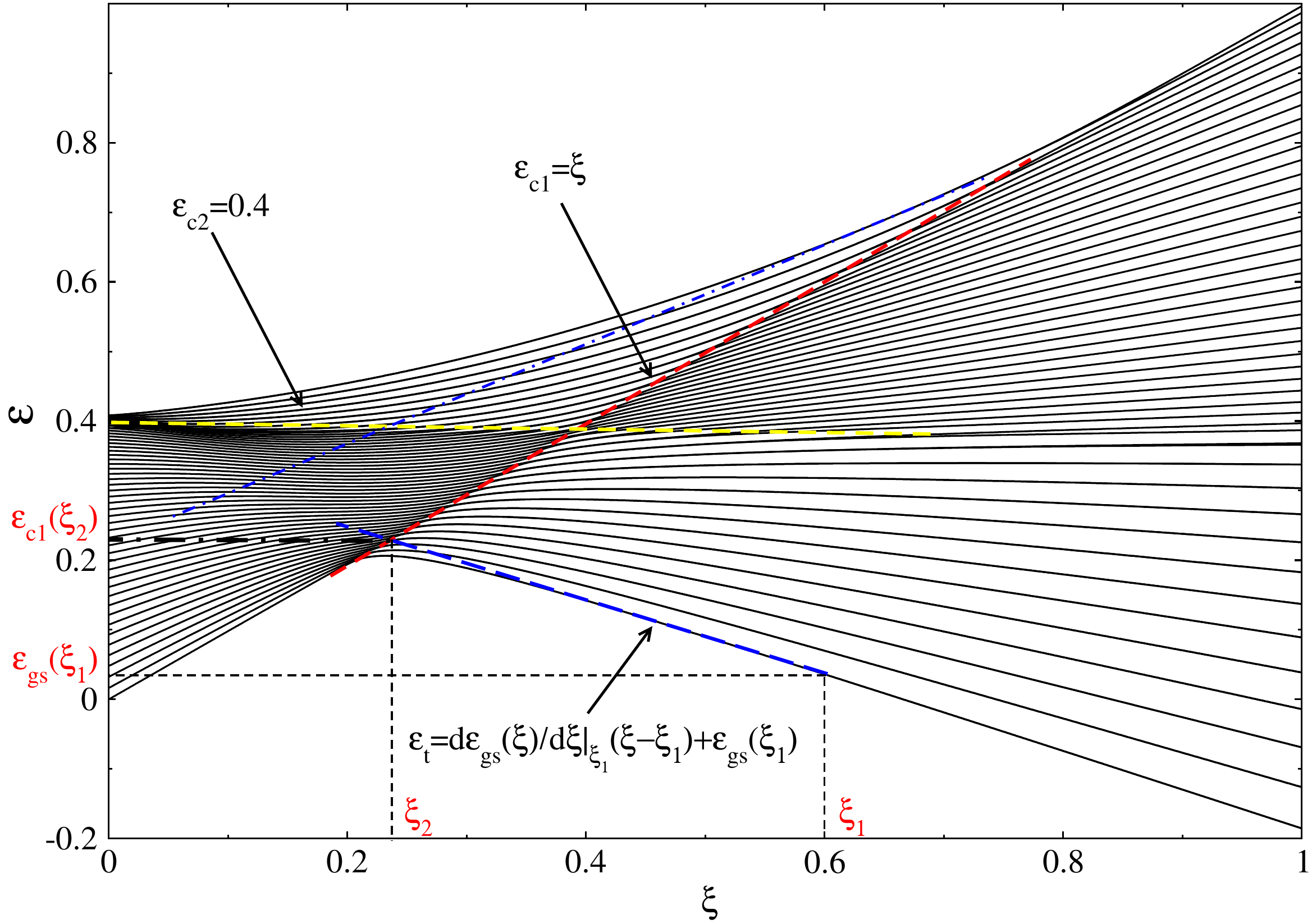}
\par\end{centering}
\caption{(Color online) Illustration of the tangent method discussed in the text for $\alpha= -0.6$. Energy spectrum of the system in the plane $\varepsilon\times\xi$ where the two ESQPT critical lines are highlighted with red and yellow dashed lines. The  dashed blue line is the tangent for the ground state curve $\varepsilon_{gs}(\xi)$ at the point $\xi_1$. This shows schematically the graphical determination of the critical quench $\xi_1 \rightarrow \xi_2$ for a given initial state. In general, the intersection of the tangent line with the critical lines provides the critical $\xi_2$ value for which the system reaches the ESQPT critical energy after the quench. The  dashed blue line stands for the tangent for the highest excited-state curve.}
\label{tangentgraph}
\end{figure*}

\begin{figure*}
\begin{centering}
\includegraphics[scale=0.65]{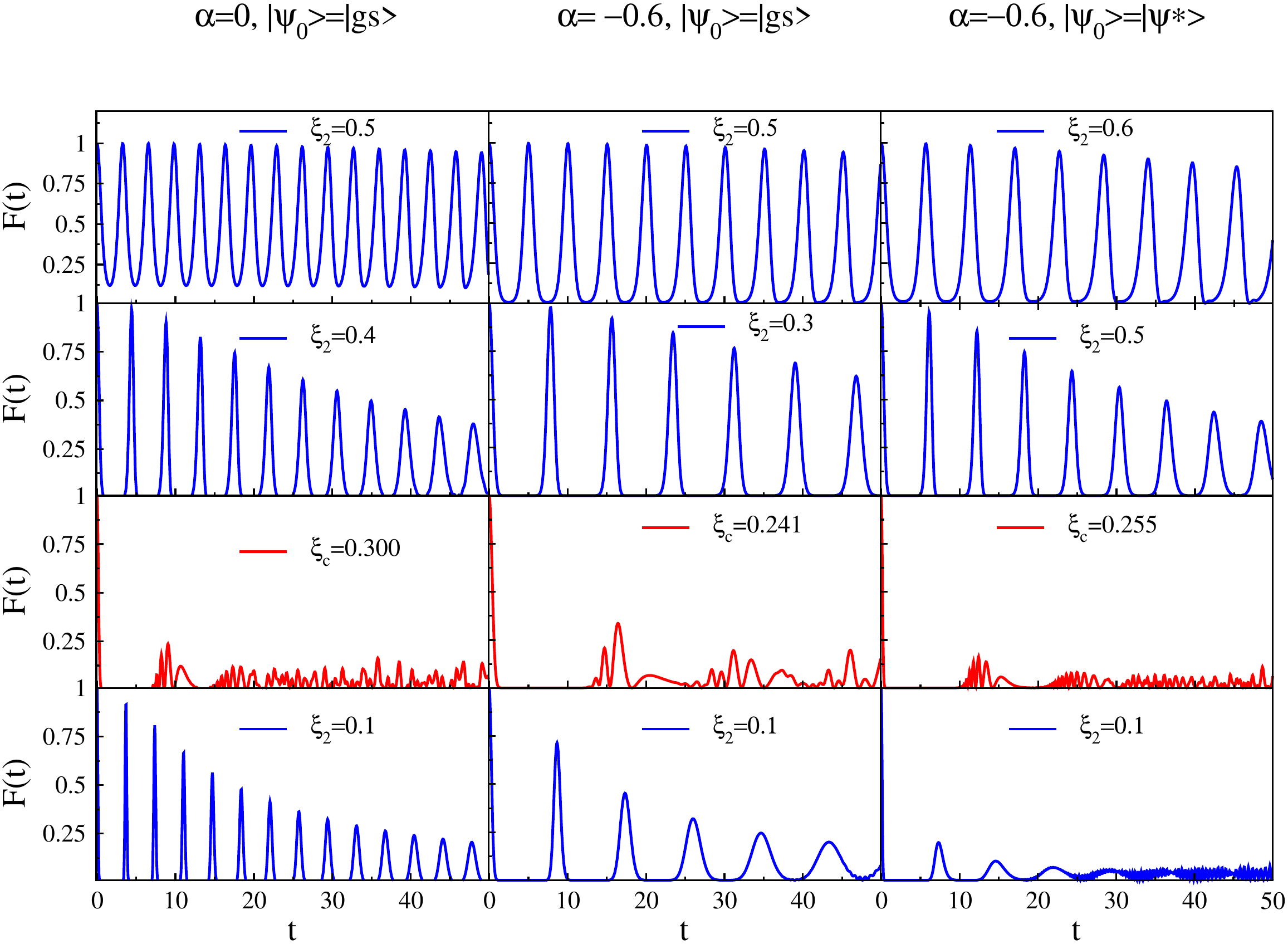}
\par\end{centering}
\caption{Survival probability $F(t)$ as a function of time ($t$) for a system  size $N=300$. The leftmost column includes $F(t)$ results for $\alpha=0$ and the middle and rightmost columns for $\alpha=-0.6$. The initial state for the leftmost and middle columns is the $\xi_1=0.6$ Hamiltonian ground state, $|\Psi_0\rangle=|gs\rangle$, and the initial state for the rightmost column is the  $\xi_1=0.7$ Hamiltonian highest excited state, $|\Psi_0\rangle=|\Psi^*\rangle$, in order to reach the second critical line of the energy spectrum. 
Different quantum quenches are shown for different values of $\xi_2$. There are some critical values for $\xi_2$, $\xi_c$, for which the system is settled in the critical energy of an ESQPT (third row) and the survival probability drops down to zero (with small random fluctuations).}
\label{fidelitygraph}
\end{figure*}

The Hamiltonian in Eq.~\eqref{eq:LMG2} depends on two control
parameters, $\xi$ and $\alpha$. In general, for negative $\alpha$
values, there exist two different ESQPTs and each one of them has a
critical energy line marked by a high level density
\cite{Gamito1, Nader2021}. Since we are interested in characterizing both ESQPTs,
a fixed value of $\alpha < 0$ is selected and the time evolution of
the system is explored after an abrupt change in the control parameter
$\xi$. The aim is to study how the system dynamics is modified by the
existence of two critical lines. In the followed quantum quench protocol, the
system is initially prepared in a certain normalized eigenstate
$|\Psi_0\rangle$ of $\hat H_1=\hat H(\xi_1)$. At time $t=0$ a quantum
quench takes place, changing $\xi$ from $\xi_1$ to $\xi_2$. Thus, the Hamiltonian for the system is now given by $\hat
H_2=\hat H(\xi_2)$ and the initial state, $|\Psi_0\rangle$, is no longer an eigenstate of $\hat H_2$ and, consequently, evolves with
time in a non trivial way.  The probability amplitude of finding the evolved state, $
|\Psi_{0}(t)\rangle$, in the initial state, $|\Psi_0\rangle$, can be
evaluated easily. The expression for this probability amplitude,
denoted as $a(t)$, is $a(t)=\langle \Psi_{0}| \Psi_{0}(t)\rangle$. The
survival probability, $F(t)$, also called nondecay probability or
fidelity, is given by the absolute square of $a(t)$,

\begin{equation}
    F(t)=|a(t)|^2= \left|\langle \Psi_{0}| \Psi_{0}(t)\rangle\right|^2 = \left|\left\langle \Psi_{0}|e^{-\imath \hat H_{2}t}|\Psi_{0}\right\rangle\right|^2 ~.
\label{Fidelity}
\end{equation}

Since our goal is to evince the effect on the system dynamics of the
external quench when reaching one of the ESQPTs' critical lines, the
determination of suitable $\xi_2$ values is very important, since the
quenched system has to reach the corresponding critical energies. This
can be achieved using the method of the tangent, developed in
Ref.~\cite{Pedro3}. In Fig.~\ref{tangentgraph}, a typical evolution of the
energy levels, $\varepsilon$, of the Hamiltonian in
Eq.~\eqref{eq:LMG2} is plotted as a function of the control parameter
$\xi$, for a value of $\alpha=-0.6$. In this figure, there is a
change in the ground state at around $\xi =0.2$ that corresponds to
the ground state QPT. In addition, two lines of high level density in
the excitation spectra are immediately apparent (separatrices, see
Ref.~\cite{Gamito1}). These lines mark the critical energy of the
ESQPTs and separate the phases in such transitions. In the case shown
in this figure, the separatrices occur at the critical energies
$\varepsilon_{c1}=\xi$ (yellow dashed line) and $\varepsilon_{c2}=0.4$ (red
dashed line). A detailed discussion on this structure, including their
dependence of the control parameters in the mean field limit, can be
found in Ref.~\cite{Gamito1}, where the static properties of the ALMG
model are presented. From Fig.~\ref{tangentgraph}, it is clear that
for analyzing the three phases one has to start from the deformed
phase $\xi > \xi_c=0.2$.  Due to the structure of our Hamiltonian,
changing $\xi$ from an initial value $\xi_1$ implies that the system
is excited along a straight line tangent to the energy line at
$\xi_1$. Thus, if the initial state is the ground state
$|\Psi_0\rangle=|gs\rangle$ for a particular $\xi_1$ value
($\xi_1>0.2$), one needs to find the value of the $\xi$ parameter,
$\xi_2$, for which the tangent of the initial energy level
$\varepsilon_{1}(\xi)$ at $\xi_1$ crosses the critical line
$\varepsilon_{c}(\xi)$ at $\xi_2$ in the plane
$\varepsilon\times\xi$. This is illustrated in Fig.~\ref{tangentgraph}
for the case in which the eigenstate $|\Psi_0\rangle=|gs\rangle$ is
the ground state of $H_1=H(\xi_1)$. It is worth noticing that, within
the range of values defined for $\xi$, using this method it is not
possible to cross both ESQPTs lines from a given initial
state. Indeed, for those values of $\xi$ above the value of the
critical $\xi_c$ for the QPT, it is only possible to reach the first
ESQPT, $\varepsilon_{c1}=\xi$, (it can be seen plotting the tangent to
the ground state line). It is worth mentioning that if one uses the
same tangent method starting from the symmetric phase ($\xi < \xi_c =
0.2$), one can reach the second ESQPT critical line,
$\varepsilon_{c2}=0.4$, (yellow line), but it would be impossible to
explore properly its impact on the dynamics of the system since one is
forced to move over the first ESQPT critical line (red dashed
line). Consequently, the tangent method from the system ground state is
suitable for the study of the first ESQPT (red dashed line), but not
the second one (yellow dashed line).

Let us first examine the $\varepsilon_{c1}=\xi$ critical line (red
dashed line), that can be reached using the tangent method from the
$\xi_1$ ground state. On the one hand, the energy of the corresponding
initial ground state is $\varepsilon_{gs}(\xi_1)$ and the equation for
the tangent line at $\xi_1$ for the curve described by the ground
state of the system in the $\varepsilon\times\xi$ plane reads
$\varepsilon_t=m (\xi-\xi_1)+\varepsilon_{gs}(\xi_1)$, where $m$ is
the slope of the tangent to the ground state curve at $\xi_1$. On the
other hand, the line of the first ESQPT (red dashed line) is
$\varepsilon_{c1}= \xi$. Therefore, both lines cross at

\begin{equation}
    \xi_2=\xi_{c1}=\frac{m ~\xi_1 - \varepsilon_{gs}(\xi_1)}{m-1}
    \label{lambdac1},
\end{equation}

\noindent where $\varepsilon_{gs}(\xi_1)=\element{gs}{\hat{H}_1}{gs}/N$
(ground state energy per particle at $\xi_1$) and the slope $m$ of the
tangent line is obtained making use of the Hellman-Feynman theorem in
Eq.~\eqref{eq:LMG2}. Indeed, $m=\left\langle
gs|\hat{H}'|gs\right\rangle/N=d\varepsilon_{gs}(\xi)/d\xi|_{\xi=\xi_{1}}$,
where $\hat{H}'=\frac{2}{S}\left(S^2-\Spin_{x}^2\right)^{2}-(S + \Spin_z )$.

A similar analysis can be performed for the 
anharmonicity-induced critical line. However, as we noticed above,
the tangent to any point along the ground state line with $\xi > \xi_c$
never crosses the second critical line (dashed yellow line) for the range of
values of $\xi$ considered in this model. Hence, to explore
this separatrix one should start from a more appropriate $\hat H_1$ eigenstate. In particular, we have selected the
highest excited state (denoted as $|\Psi^*\rangle$). As with the ground state, the most excited state of our system is well-defined in the thermodynamic limit by a coherent-state \cite{Lerma2018}. Then, our initial
state is now $|\Psi_0\rangle=|\Psi^*\rangle$ of $\hat H_1$. Let us denote the slope of the tangent to the energy line of the highest
state at $\xi_1$ as
$m_2$. Then, this tangent line will reach the anharmonicity-induced ESQPT
line given by $\varepsilon_{c2}=\varepsilon_{0}$ which is a
constant. In the $\alpha = -0.6$, the value of $\varepsilon_{0}=0.4$ was computed with a mean
field formalism \cite{Gamito1}. Therefore, the value for the critical
$\xi$, $\xi_{c2}$, reads

\begin{equation}
    \xi_{c2}=\frac{m_2~\xi_1+\varepsilon_{0}-\varepsilon_{\Psi^*}(\xi_1)}{m_2}
    \label{lambdac2},
\end{equation}

\noindent where $\varepsilon_{\Psi^*}(\xi_1)=\element{\Psi^*}{\hat{H}_1}{\Psi^*}/N$ and $m_2=\left\langle \Psi^*|\hat{H}'|\Psi^*\right\rangle/N=d\varepsilon_{\Psi^*}(\xi)/d\xi|_{\xi=\xi_{1}}$ is the slope of the corresponding tangent line.

\medskip

Once a way of crossing both ESQPT lines is available, the dynamic
evolution of the system and the effect of crossing an ESQPT line can
be examined. This can be accomplished computing the survival
probability $F(t)$ Eq.~(\ref{Fidelity}).  Results for $F(t)$ as a
function of time are shown in Fig.~\ref{fidelitygraph} for $N=300$,
$\alpha = 0$ (left column) and $-0.6$ (center and right columns) and
different initial states (either the ground state
$|\Psi_0\rangle=|gs\rangle $ in the left and central columns or the
most excited state $|\Psi_0\rangle=|\Psi^*\rangle$ in the right
column) for selected $\xi-$values. The $\alpha=0$ case  in the leftmost
panels is included for the sake of completeness and reference. The panels in this column depict the time evolution of the survival
probability for decreasing values of $\xi_2$, starting always from the
ground state $\ket{gs}$ for $\xi_1=0.6$. The calculated $\xi_2$ at the
crossing with the ESQPT is $\xi_c=0.3$. In general, the survival
probability has a regular oscillatory behaviour except in the region
close to the ESQPT critical energy, $\xi_2=0.3$,
where the system undergoes an ESQPT and the survival probability
suddenly drops down to zero and starts to oscillate randomly with
small amplitudes. Once the critical energy for the ESQPT is crossed,
the survival probability starts to oscillate in a regular way
again. This phenomenon was reported for the first time in
Ref.~\cite{Pedro1}. In the central and rightmost columns the same observable is plotted
including a non-zero anharmonic term ($\alpha = -0.6$).
%Second column corresponds to use as initial state the ground state $|\Psi_0\rangle$ for $\lambda_1=0.6$ and in third column the initial state is the highest energy state $|\Psi^*\rangle$ for $\lambda_1=0.7$. 

%%%%%%%%%%%%%%%%%%%%%%%%%%%%%%%%%%%%%%%%%%%%%%%%%%%%%%%%%%%%%%%%%%%%%%%%%%%%%%%%%%%%%%%%%%%%%%%%%%%%%%%%%%%%%%%%%

%%%%%%%%%%%%%%%%%%%%%%%%%%%%%%%%%%%%%%%%%%%%%
%%%%%%%%%%%%%%%%%%%%%%%%%%%%%%%%%%%%%%%%%%%%

In the panels of the second column of Fig.~\ref{fidelitygraph}, the
survival probability is depicted for decreasing values of $\xi_2$ and
starting always from the ground state of a Hamiltonian with
$\xi_1=0.6$ and $\alpha=-0.6$. Due to the negative $\alpha$ value, the
system undergoes two ESQPTs, displayed in the spectrum by means of
critical lines with a noteworthy accumulation of energy levels
(see Fig.~\ref{tangentgraph}). One of the two critical lines (red
dashed line) can be traced back to the ESQPT already present in the
$\alpha=0$ case \cite{ESQPT}.  However, the second one (yellow dashed
line) is linked to the presence of the anharmonic term in the
Hamiltonian \cite{Gamito1}. The nature and physical interpretation of the
anharmonicity-induced ESQPT is different from the already known ESQPT
associated with the ground state QPT. Hence, in principle, there is no
{\it a-priori} reason for both behaving in the same way. However, as
we see if we compare the results for the critical $\xi_c$ values in the
different columns, the results obtained for the $\alpha=0$ and the
anharmonic cases are similar. The survival probability is
oscillatory and regular except once $\xi_2$ is close to $\xi_c$, the critical
value for the first or second ESQPT.
%% , that in this case is calculated to be at $\xi_c=0.241$
In all cases, when $\xi_2 = \xi_c$, the quenched system reaches the critical energy and the survival probability suddenly drops down to
zero and oscillates randomly with a small amplitude (red curves). This is similar to what happens in the $\alpha=0$ case. Once $\xi_2$ is
smaller than $\xi_c$, a periodic oscillatory decaying behaviour is
observed in $F(t)$. As explained
above, the quench from the ground state never reaches the second ESQPT
line. For that purpose, one has to start from a different initial
state. Thus, in order to explore how $F(t)$ is affected by the second
ESQPT, the quantum quench is performed using as an initial state the
highest excited state of the system,  $|\Psi^*\rangle$, for a given value
of $\xi_1$. In this way, the second critical line (yellow dashed line)
for the ESQPT is accessible after the quench. In the panels of the right column of
Fig.~\ref{fidelitygraph}, the survival probability for decreasing
$\xi_2$ values is plotted for an initial state equal to the highest excited
state of the Hamiltonian with $\xi_1=0.7$ and $\alpha=-0.6$. For this parameter selection,
the second critical line is reached at $\xi_c=0.255$. In this column,
again, results are very similar to the ones obtained in the preceding cases. The fidelity $F(t)$
oscillates regularly while $\xi_2>\xi_{c2}$, but
when the $\xi_2$ parameter gets close to the critical value,
$\xi_{c2}=0.255$, the survival probability drops down to zero and
randomly oscillates with a small amplitude. Once the critical line is
crossed, $F(t)$ recovers an oscillatory decaying periodic
behavior, but at a certain time, this periodic oscillatory behaviour
becomes distorted. The reason for this phenomenon is that the tangent
line to the highest excited state curve at $\xi_1$ in the plane
$\varepsilon\times\xi$ remains very close to the critical line
$\varepsilon_{c2} = 0.4$ after the quench for lower values of $\xi_2$
up to 0. One should note that when starting from the highest excited
state the first ESQPT critical line is not accessible after the quench
(see Fig.~\ref{tangentgraph}).

\begin{figure}
\begin{centering}
\includegraphics[scale=0.25,angle=0]{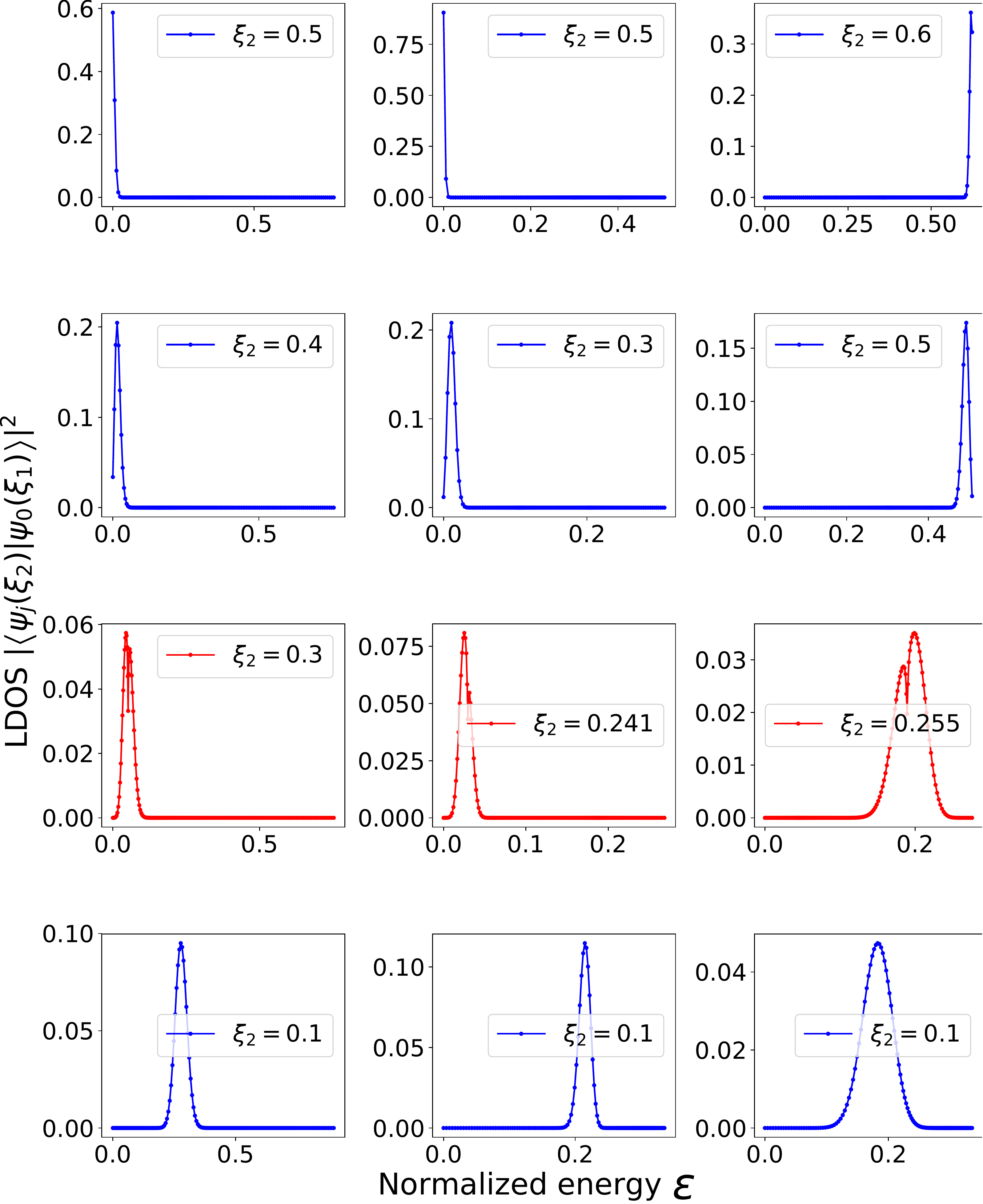}
\par\end{centering}
\caption{LDOS $\left|\braket{\psi_j(\xi_2)}{\psi_0}\right|^2$ as a function of the normalized excitation energy $\varepsilon_j$ (arb. units) for systems with $\xi_1=0.6$ and $\alpha=0.0$ (left column),  $\xi_1=0.6$ and $\alpha=-0.6$ (middle column), and $\xi_1=0.7$ and $\alpha=-0.6$ (right column) ($N=300$ in all cases). The chosen states are the ground state $\ket{\psi_0}=\ket{\text{gs}(\xi_1)}$ (left and middle columns) and the most excited state with even parity $\ket{\psi_0}=\ket{\Psi^*(\xi_1)}$ (right column), expressed in all cases in the basis of eigenstates of the Hamiltonian $\hat{H}(\xi_2,\alpha)$, being $\xi_2$ the quench parameter. The cases that correspond to a critical value of $\xi_2$ (third row) are plotted using red color.}
\label{ldosgraph}
\end{figure}

%\textcolor{red}{Include strength function of LDOS definition and that the survival probability is the Fourier transform of this quantity. Include LDOS figure correctly labeled and explain the $F(t)$ patterns accordingly.}

If we denote the eigenstates of $\hat{H}(\xi_i,\alpha)$ with $i=1,2$ as $\ket{\psi_j(\xi_i)}$ with $j = 0, 1, \ldots, N/2$, then we can write the initial state  $\ket{\Psi_0}$ in the basis of $\hat H_2= \hat{H}(\xi_2,\alpha)$ eigenfunctions as $\ket{\Psi_0} = \sum_j C_j\ket{\psi_j(\xi_2)}$ and then
\begin{align}
    F(t) &= \left|\left\langle \Psi_{0}|e^{-\imath \hat H_{2}t}|\Psi_{0}\right\rangle\right|^2 \nonumber\\
    &= \left|\sum_j\left|C_j\right|^2 e^{-iE_jt}\right|^2 =  \left|\int dE e^{-iEt} \rho_0(E) \right|^2~, \label{ldos}
\end{align}
\noindent where $E_j$ is the energy of the $j$-th  $\hat H_2$ eigenstate and $\rho_0(E) = \sum_j \left|C_j\right|^2 \delta(E-E_j)$, called the strength function or local density of states (LDOS) \cite{Santos2016, Borgonovi2016}, is the energy distribution of $\ket{\Psi_0}$ weighted by the $C_j$ components. 

From Eq.~\eqref{ldos} it is clear that the fidelity $F(t)$ is the absolute value of the LDOS Fourier transform squared and this quantity can provide some clues on the $F(t)$ time dependence for the quench at the critical values $\xi_c$, denoted in red in the third row of Fig.~\ref{fidelitygraph}. In Fig.~\ref{ldosgraph} we plot the LDOS for the same cases included in Fig.~\ref{fidelitygraph}, hence in the first column, we show the LDOS for the ground state of $\hat{H}_1 = \hat{H}(\xi_1 = 0.6,\alpha = 0)$ for different $\hat{H}_2$ cases, all of them with $\alpha = 0$. In the second and third columns we depict the LDOS for initial states that are the ground state of  $\hat{H}_1 = \hat{H}(\xi_1 = 0.6,\alpha = -0.6)$ and the most excited state of $\hat{H}_1 = \hat{H}(\xi_1 = 0.7,\alpha = -0.6)$. The LDOS for the critical quench values are depicted in red. It can be clearly seen that, for all columns, in the critical control parameter cases the LDOS is nonzero at the ESQPT critical energy and has a clear local minimum at this energy value. The Fourier transform of such LDOS produces the particular time dependence shown in the panels of the  Fig.~\ref{fidelitygraph} third row.

\medskip

Another quantity of interest, inspired on Loschmidt's objections to Boltzmann H theorem, is the Loschmidt echo  \cite{Peres1984,Jalabert2001}. This quantity, considered a probe to the sensibility of a system dynamics under perturbations, is used to benchmark the reliability of quantum processes \cite{Gorin2006}. It was shown to be a valid QPT detector \cite{Quan2006} and, more recently, it has been used to check the influence of the ESQPT on the dynamics of the LMG model \cite{LE_QWang_2017}. Consider an initial wave function, $\ket{\psi}$, which evolves a time $t$ under a Hamiltonian $\hat{H}_1$, $\ket{\psi(t)} = e^{-i\hat{H}_1 t}\ket{\psi}$. We can reverse the time evolution with another Hamiltonian $\hat{H}_2$, $e^{i\hat{H}_2 t}e^{-i\hat{H}_1 t}\ket{\psi}$. The squared overlap of the resultant state with the initial state $\ket{\psi}$ is the Loschmidt echo (LE), denoted as $M(t)$ \cite{Jalabert2001, LE2012},
\begin{equation}\label{LEcho}
    M(t)=\left|\bra{\psi}e^{i\hat{H}_2 t}e^{-i\hat{H}_1 t}\ket{\psi}\right|^2~.
\end{equation}
Another physical interpretation of this quantity is possible, since Eq.~\eqref{LEcho} is the distance between the same initial state once it is evolved for a time $t$ with two different Hamiltonian operators. One of the properties of ground state and excited state QPTs is that, near the critical region, states are quite sensitive to perturbations. A way to quantify this effect is computing the LE for the eigenstates of the system $\hat{H}_1=\hat{H}(\xi,\alpha)$ with a time-reversal under $\hat{H}_2=\hat{H}(\xi+\delta,\alpha)$,
\begin{equation}
    M_j(t)= \left|\left\langle \psi_j(\xi,\alpha)\left| e^{i\hat{H}(\xi+\delta,\alpha)t}  \right|  \psi_j(\xi,\alpha)\right\rangle\right|^2~,
\end{equation}
where $\ket{\psi_j(\xi, \alpha)}$ is the $j$-th eigenstate  of $\hat{H}_1$ and $\delta$ is a small perturbation. The LE, as well as its long-time average value, detects the ESQPT in the LMG model without anharmonicity \cite{LE_QWang_2017}. 

\begin{figure}
    \centering
    \includegraphics[width=0.5\textwidth]{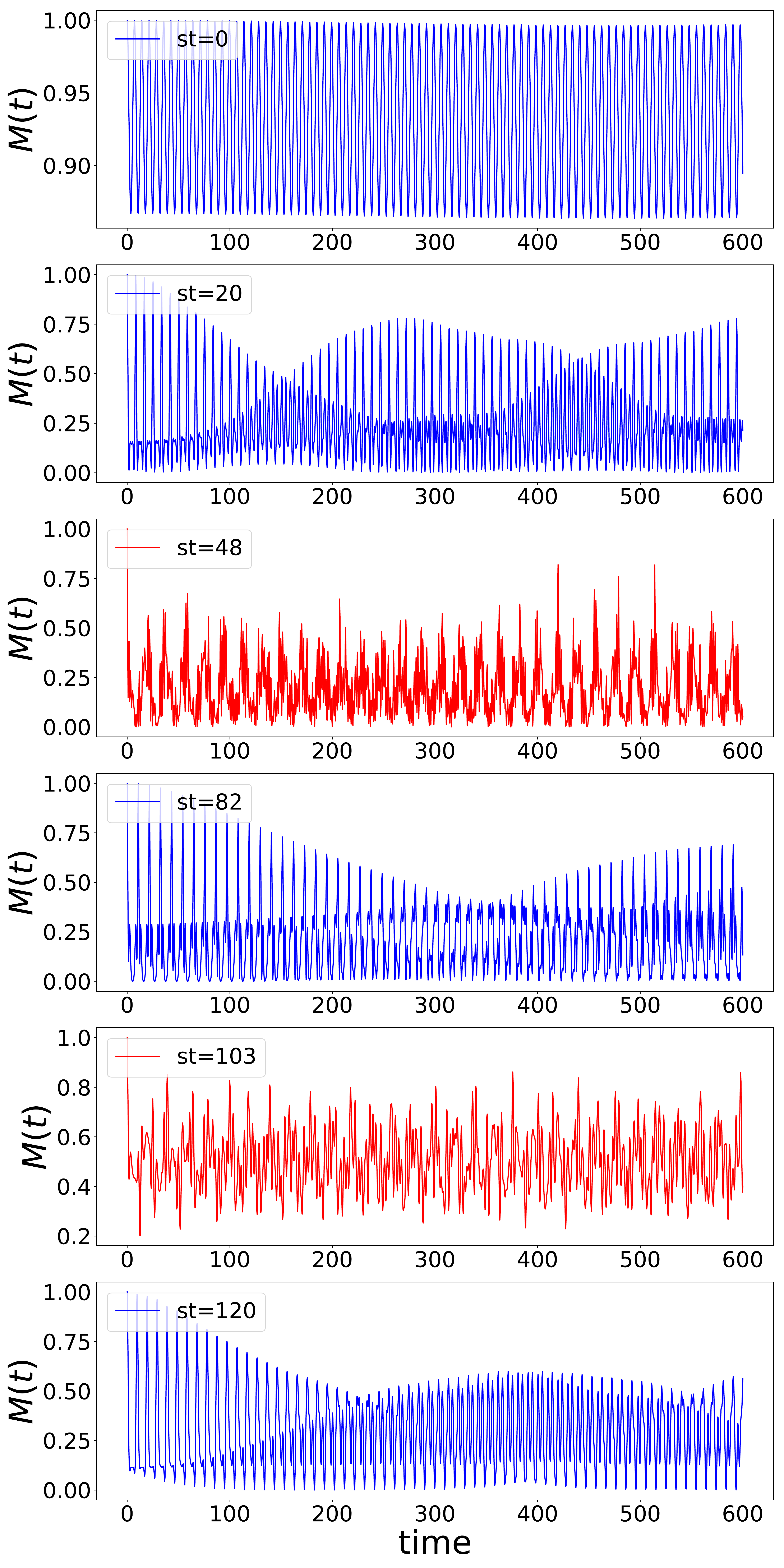}
    \caption{Loschmidt echoes for a system with $\xi=0.3$, $\alpha=-0.6$, $N=300$, and a perturbation across the control parameter $\xi$ of $\delta=0.01$. From top to bottom we display $M_j(t)$ for the $j$-th state with even parity: $0$, $20$, $48$, $82$, $103$ and $120$. The states closer to the ESQPTs are plotted in red.}
    \label{fig:LE_t}
\end{figure}

In Fig.~\ref{fig:LE_t} we plot $M_j(t)$ for several eigenstates of a system with $\xi=0.3$ and $\alpha=-0.6$. The total number of bosons is $N=300$, the system has been perturbed with $\delta=0.01$, and only states with even parity are considered. Results are shown for $j=0, ~20, ~48, ~82, ~103$, and $120$. The two states with energies closest to ESQPTs critical energies ($j=48$ and $103$) are plotted in red. As expected, the ground state $j=0$ perform small oscillations with a single frequency around a value close to one, with a maximum value equal to one. Other states far from the critical region, as $j=20, ~82$, and $120$, have a more complex oscillation pattern, not harmonic, with a larger amplitude and without reaching unity in the considered time range. However, in the case of eigenstates close to the critical energy, $j=48$ and $103$, $M(t)$ is only one for $t=0$ and the oscillations of the LE are of a much more irregular nature, something similar to what happens for the fidelity $F(t)$ in Fig.~\ref{fidelitygraph}. %This issue turns into a high dispersion for states close to the ESQPTs.

\begin{figure}
    \centering
    \includegraphics[width=0.5\textwidth]{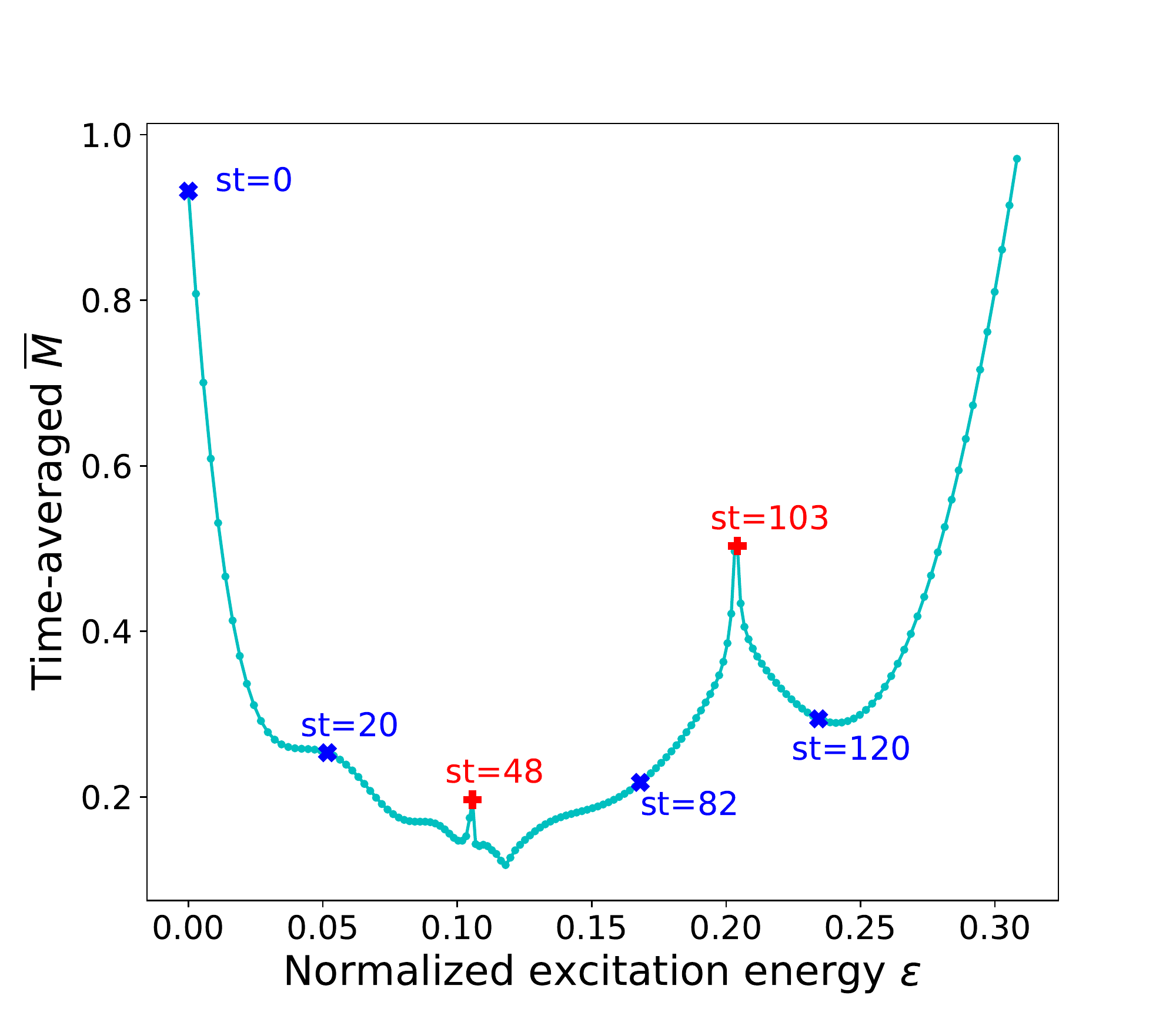}
    \caption{Time-averaged of $M(t)$ versus the normalized excitation energy $\varepsilon$ for the same system introduced in Fig.~\ref{fig:LE_t}. The highlighted states (red plus symbols for transition states and blue crosses for others) correspond to the states studied in Fig.~\ref{fig:LE_t}.}
    \label{fig:LE_prod}
\end{figure}

As shown in Ref.~\cite{LE_QWang_2017}, the time averaged value of the LE for the $j$-th eienstate, $\overline{M}_j$, is a convenient probe to detect an ESQPT. This quantity is defined as
\begin{equation}
    \overline{M}_j=\lim_{T\to\infty}\frac{1}{T}\int_0^T dt~M(t)= \sum_k \abs{c_{jk}^{\delta}}^4~,
\end{equation}
where $c_{jk}^{\delta}$ are the coefficients of the $j$-th eigenfunction of the Hamiltonian operator $\hat{H}(\xi+\delta,\alpha)$, expressed in the basis of eigenstates of $\hat{H}(\xi,\alpha)$: $\ket{\psi_j(\xi+\delta,\alpha)}=\sum_k  c_{jk}^{\delta} \ket{\psi_k(\xi,\alpha)}$. The LE time averaged value is equal to the inverse of the participation ratio (PR) \footnote{The PR is defined as follows $P(\psi) = \frac{1}{\sum_m |a_m|^4}$, where $a_m$ are the components of $\ket{\psi}$ in a given basis.} of $\ket{\psi_j(\xi+\delta,\alpha)}$ computed using the basis $\left\{\ket{\psi_k(\xi,\alpha)}\right\}$. In Fig.~\ref{fig:LE_prod} we plot the time averaged LE versus the normalized excitation energy for all even parity states of the system studied in Fig.~\ref{fig:LE_t}. The states included in Fig.~\ref{fig:LE_t} have been marked using red pluses for critical ones ($j=48$ and $103$) and blue crosses for others ($j=20,~82$, and $120$). $\overline{M}_j$ has local maxima located for the eigenstates close to the critical energies, as it was observed in the LMG model without anharmonicity \cite{LE_QWang_2017}. Hence, the time averaged LE  detects the new ESQPT associated to the anharmonic term in the LMG Hamiltonian and confirms that this quantity is a good ESQPT probe. 
%%%%%%%%%%%%%%%%%%%%%%%%%%%%%%%%%%%%%%%%%%%%%%%%%%%%%%%%%%%%%%%%%%%%%%%%%%%%%%%%%%%%%%%%%%%%%%%%%%%%%%%%%%%%%%%%%%%%%%%%%%%%%%%1
\section{ESQPTs and OTOC}

Out-of-time-order correlators (OTOCs), that appeared for the first
time in the '60s in the context of superconductivity \cite{Larkin1969}, are
a four-point temporal correlation function able to measure the
entanglement spread in a quantum system from the degree of
noncommutativity in time between operators. Since then, after a long
period of relative inactivity, there has been a tremendous frenzy
around this concept on various fronts \cite{Swingle18}. They returned
to the limelight with the proposal of OTOCs as a viable quantum chaos
indicator, due to its exponential increase at early times in certain
systems \cite{Shenker2014, kitaev, Roberts2015, Maldacena2016}, and to
diagnose the scrambling of quantum information \cite{Swingle2016a,
  Lewis2019, Xu2019, Niknam2020}. Besides, OTOCs are sensitive probes
for quantum phase transitions
\cite{Shen2017,Heyl2018,CurroOTOC,Ceren2019, Nie2020,
  Lewis2020, KRivera2023}. Despite the fact that the experimental access to
out-of-time-order correlators is hindered by the unusual time ordering
of its constituents operators that precludes the measurement using
local operators, several approaches using different experimental
platforms have successfully provided OTOC results \cite{Li2017,
  Garttner2017, Wei2018, Landsman2019, Pegahan2021, Green2022,
  Braumuller2022, Li2022}.

Given two operators, $\hat{W}$ and $\hat{V}$, it is possible to probe
the spread of $\hat{W}(t)$ with $\hat{V}$ through the expectation
value of the square commutator

\begin{equation}
 {\cal C}_{w,v}(t)=\left\langle \left[\hat{W}(t),\hat{V}(0)\right]^{\dagger}\left[\hat{W}(t),\hat{V}(0)\right]\right\rangle,
\label{C}
\end{equation}

\noindent where
$\hat{W}(t)=e^{\imath\hat{H}t}\hat{W}e^{-\imath\hat{H}t}$ is the
operator $\hat{W}$ in the Heisenberg's representation \cite{Swingle18,
  Swingle2016b, Hashimoto17, Hashimoto20, Akutagawa20}. The
expectation value is usually computed in the canonical ensemble.
However, in recent works, it has also been computed over given initial
states or over the system eigenstates (microcanonical OTOC)
\cite{Hashimoto17, Hashimoto20}. The squared commutator Eq.~(\ref{C})
can be rewritten as ${\cal C}_{w,v}(t)= {\cal A}_{w,v}(t) -2{\cal F}_{w,v}(t)$. The
first term is a two-point correlator,
${\cal A}_{w,v}(t) = \left\langle
  \hat{W}^{\dagger}(t)\hat{V}^{\dagger}(0)\hat{V}(0)\hat{W}(t)\right\rangle
+ \left\langle
  \hat{V}^{\dagger}(0)\hat{W}^{\dagger}(t)\hat{W}(t)\hat{V}(0)\right\rangle$
and the out-of-time order appears in ${\cal F}_{w,v}(t)$, the real part of a
four-point correlator
\begin{equation}
{\cal F}_{w,v}(t)=\Re\left[\left\langle \hat{W}^{\dagger}(t)\hat{V}^{\dagger}(0)\hat{W}(t)\hat{V}(0)\right\rangle\right].
\label{OTOC}
\end{equation}
Without loss of generality, if we consider operators that are unitary,
then Eq.~\eqref{C} reads ${\cal C}_{w,v}(t)=2-2{\cal F}_{w,v}(t)$.

In a recent LMG model study, the ESQPT effects on the microcanonical
OTOC and the OTOC following a quantum quench were explored for
$\hat{W}=\hat{V}=\Spin_{x}/S$~\cite{CurroOTOC}. The time evolution of
the OTOC after a sudden quench was analyzed and it was concluded that
the equilibrium value (the long time average value) of this observable
can be used as a good marker for the ESQPT because it behaves as an
order parameter, able to distinguish between the phases below and
above the ESQPT, respectively. Our goal, here, is to analyze how the
OTOC behaves once the ALMG system goes through the
anharmonicity-induced ESQPT line. This study is of relevance since the
physical nature of this ESQPT is different from the one of the already
known ESQPT for the usual LMG model. Moreover, the possibility of
using an OTOC as an order parameter for both ESQPTs is
considered. %% Through this work, we consider operators such as $\left[\hat{W}(0),\hat{V}(0)\right]=0$. Actually, this is not a very restrictive condition and alleviates calculations.

\medskip

We have used in our analysis the microcanonical OTOC
\cite{Hashimoto17, ChavezC19}, defined as

\begin{equation}
{\cal F}_{n}(t)=\Re\left[\element{n}{\hat{W}^{\dagger}(t)\hat{V}^{\dagger}(0)\hat{W}(t)\hat{V}(0)}{n}\right]~,
\label{OTOCcanonical}
\end{equation}
\noindent where  the state $\ket{n}$  is the $n$-th eigenstate  of the
Hamiltonian Eq.~\eqref{eq:LMG2}, whose energy  is $E_n$. This state is
computed  for  a  given  set  of  Hamiltonian  parameters,  $\xi$  and
$\alpha$.

Following Ref.~\cite{CurroOTOC}, we have first selected
$\hat{W}=\hat{V}=\Spin_{x}/S$ as the OTOC operators. The reason behind
this election is twofold. On the first hand, the expectation value of
the $\Spin_{x}$ operator is known to be an order parameter for the QPT
in the LMG model, and it has also been shown in previous works that
it behaves as an order parameter for the ESQPT \cite{Pedro4}. On the second hand, the $\Spin_{x}$ operator is related with
the breaking of parity symmetry in the spectrum eigenstates
\cite{Santos2016}. However,
the obtained results (not shown) indicate that in this case the
$F_{n}(t)$ equilibrium value only detects the occurrence of the first
ESQPT, independently of its nature, and not the second one. We decided
to explore other possibilities such as $\hat{W}=\Spin_{y}/S$,
$\hat{V}=\Spin_{x}/S$ or $\hat{W}=\Spin_{+}/S$,
$\hat{V}=\Spin_{-}/S$. In both cases we obtain the expected results,
with equilibrium values sensitive to the anharmonicity-induced ESQPT
in the symmetric phase and to the two ESQPTs in the broken symmetry
phase.

\begin{figure*}
\begin{centering}
  \includegraphics[scale=0.55]{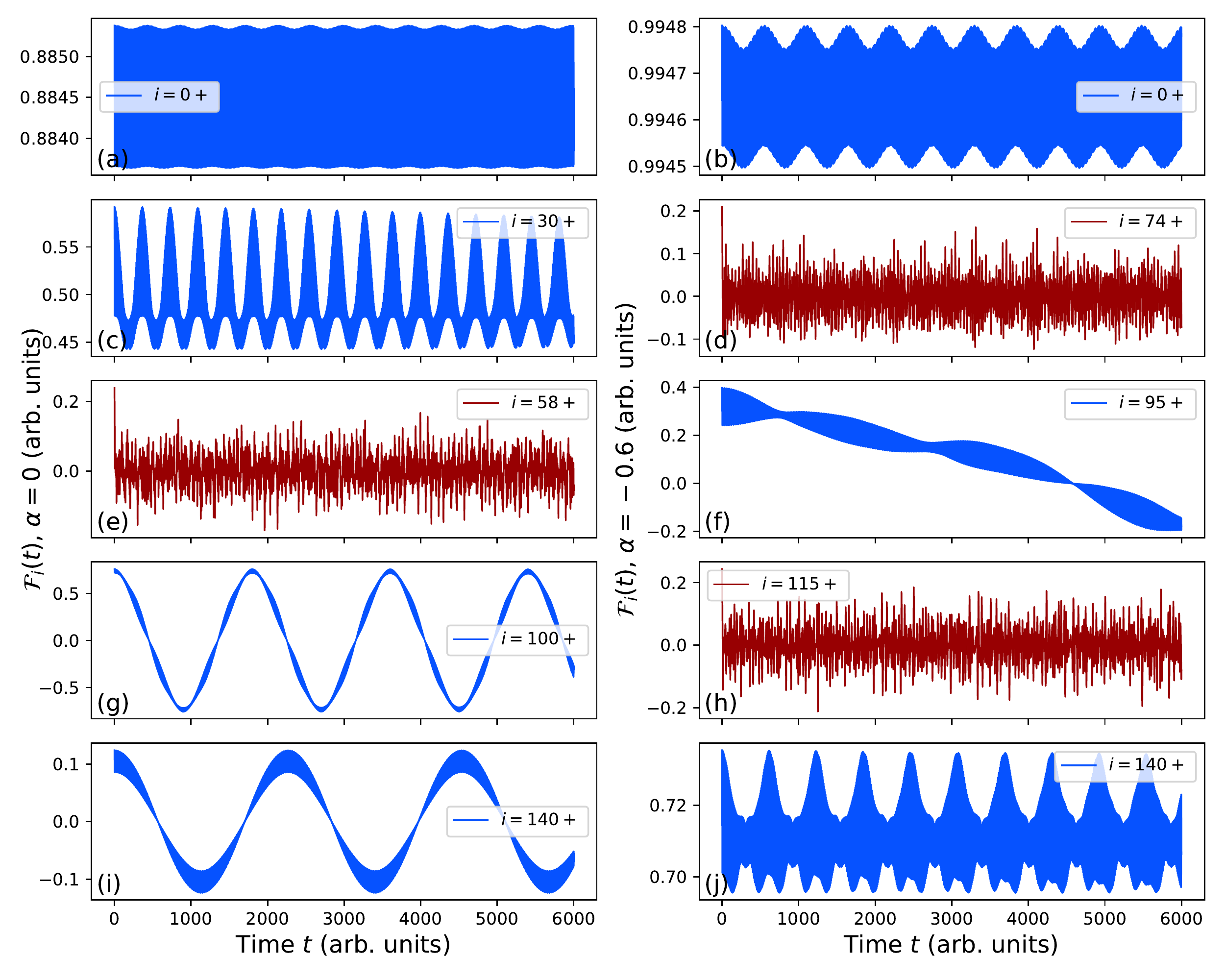}
\par\end{centering}
\caption{(Color online) Time evolution of the microcanonical OTOC,
  ${\cal F}_{i}(t)$, for selected positive parity eigenstates of an ALMG
  model with a system size $N=300$. In all panels $\xi=0.5$, the left
  column panels refers to ${\cal F}_{i}(t)$ for $\alpha=0$ and the right
  column panels include results for $\alpha=-0.6$. OTOCs for different
  initial states are shown: for the left column from top to bottom;
  (a) $\ket{i=0}$ (ground state), (c) $\ket{i=30}$, (e) $\ket{i=58}$,
  (g) $\ket{i=100}$, and (e) $\ket{i=140}$. For the right column from
  top to bottom: (b) $\ket{i=0}$ (ground state), (d) $\ket{i=74}$, (f)
  $\ket{i=95}$, (h) $\ket{i=115}$, (j) $\ket{i=140}$. There are some
  energies in which the eigenstate is settled at the critical energy
  of an ESQPT. These are the cases for panels (e) in the left column
  and (d) and (h) in the right column, highlighted using a red color.}
\label{otocgraph}
\end{figure*}

%%%%%%%%%%%%%%%%%%%%%%%%%%%%%%%%%%%%%%%%%%%%%%%%%%%%%%%%%%%%%%%%%%%%%%%%%%%%%%%%%%%%%%%%%%%%%%0.04277478, 0.39689748, 0.44550815, 0.59895678, 0.64589824

Numerical solutions for the time evolution of the OTOC
Eq.~\eqref{OTOCcanonical} with $\hat{V}=\Spin_{-}/S$ and
$\hat{W}=\Spin_{+}/S$ are presented in Fig.~\ref{otocgraph}. These are
results for a selected set of positive parity states of a system with
size $N=300$ that are obtained by the diagonalization of the
Hamiltonian Eq.~\eqref{eq:LMG2}. The time evolution of the
microcanonical OTOC is depicted for different initial states and
$\xi=0.5$ with either $\alpha=0$ (left-column panels) or $\alpha=-0.6$
(right-column panels). Despite the different operators included in the OTOC, a
quite similar phenomenology to that pointed out in
Ref.~\cite{CurroOTOC} is observed. However, it is worth to emphasize
that if we kept $\hat{V}=\hat{W}=\Spin_{x}/S$, once the first critical
energy is crossed, the time average value of the OTOC is zero as
${\cal F}_{n}(t)$ oscillates around
zero. %% Indeed, in both columns regardless the initial state, the imaginary part of $F_{n}(t)$ (Eq. \eqref{OTOCcanonical}) is always an oscillating function around zero.

The behaviour of the microcanonical OTOC, ${\cal F}_{i}(t)$, depends on the
region of the spectrum in which the system is located. Particularly,
${\cal F}_{i}(t)$ develops a regular behavior, with small amplitude
oscillations around a positive value. This value decreases until the
${\cal F}_{i}(t)$ oscillates around zero, when the critical energy value is
reached. For the states close to the ESQPT critical energy (red color
curves), not only ${\cal F}_{i}(t)$ oscillates around zero, but it also
behaves in a highly irregular way, as in Ref.~\cite{CurroOTOC}. This
is a feature shared by both columns in Fig.~\ref{otocgraph}, though in
the right column panels the second and fourth panel correspond to
critical energies for the two ESQPTs that arise in this case.

% Instead of unity, as in Ref.~\cite{CurroOTOC}, this value is close to
% zero when the system has an energy close to the ground state
% energy. The microcanonical OTOC oscillates around larger values until
% it reaches a maximum and then, this value decreases until the
% $F_{i}(t)$ oscillates around zero again, when the critical energy
% value is reached. For the states closest to the ESQPT critical energy
% (red color curve), not only $F_{i}(t)$ oscillates around zero, but it
% also behaves in a highly irregular way, as in
% Ref.~\cite{CurroOTOC}. This is a feature shared by both columns in
% Fig.~\ref{otocgraph}, though in the right column panels the second and
% fourth panel correspond to critical energies for the two ESQPTs that
% arise in this case.

\medskip

%%%%%%%%%%%%%%%%%%%%%%%%%%%%%%%%%%%%%%%%%%%%%%
Let us now to discuss in more detail the left column
($\alpha=0$). Remind that in this case there is just one ESQPT located
in the mean field limit at energy $\varepsilon=\xi$, its value for
these plots is $\varepsilon=\xi=0.5$. We have selected the ground state and four other positive
parity eigenstates, $i = 0$, $30$, $58$, $100$, and $140$ in panels (a), (c), (e), (g), and (i), respectively.
The state with the closer energy to the critical ESQPT energy is $i = 58$ --panel (e)-- where the ESQPT precursors are clearly manifested. In
the cases with energies below the critical energy
${\cal F}_{i}(t)>0$. However, as can be seen in panel (e), once the critical
energy for the ESQPT is reached, the OTOC oscillates randomly around
zero. For energies larger than the critical energy --left panels (g) and
(i)-- the ${\cal F}_{n}(t)$ display high and low frequency oscillations around a zero mean value. Therefore, the steady-state
value of ${\cal F}_{n}(t)$ will be equal to zero for these states. As one
goes up in energy in the spectrum, the same kind of oscillatory
behavior is observed, with smaller amplitudes.

\medskip

%%%%%%%%%%%%%%%%%%%%%%%%%%%%%%%%%%%%%%
There are some new features arising in the right column panels, that include ${\cal F}_{i}(t)$ results for the  $\alpha=-0.6$ anharmonic case. As previously mentioned, in this case there are two critical ESQPT lines that in the mean field limit lie at $\varepsilon=1+\alpha=0.4$ and $\varepsilon=\xi=0.5$. We show the results for the two eigenstates with local minimum PR values $i=74$ and  
$115$ in panels (d) and (h) of the right column in Fig.~\ref{otocgraph}. Again, the  ${\cal F}_{i}(t)$ OTOC oscillations at the
critical lines are markedly irregular. These two states have been highlighted using red color. The
other three values included in the right column of
Fig.~\ref{otocgraph} are $i=0$ (ground state), $95$, and
$140$. In the region between the two critical lines, the
envelope for ${\cal F}_{i}(t)$ has a sine-like oscillatory behaviour around
zero, so its steady-state value equals zero. Once the second critical line is crossed and the system energy increases, ${\cal F}_{i}(t)$ presents again
an oscillatory behaviour around positive values, as can be clearly seen in 
Fig.~\ref{otocgraph}. It is worth pointing out that the characteristic times of the different microcanonical OTOCs span a wide range of frequencies. In particular, panel (f) in Fig.~\ref{otocgraph} exhibits a much longer period (smaller frequency) than the rest of the panels. The oscillatory frequency of the four-point correlator can be traced back to energy differences between pairs of states of different parity \cite{KRivera2023}. Therefore, whenever different parity eigenstates are degenerate, the stationary value of the OTOC has a non-zero contribution. This occurs at energies less than the critical energy of the first ESQPT and above the critical energy of the second ESQPT. The OTOC associated with eigenstates whose energies 
are either just under the critical energy of the first ESQPT or above the critical energy of the second ESQPT have small frequencies due to the small energy differences because the degeneracy starts splitting. A similar small-frequency OTOC can be observed for energies in between both ESQPTs ---as shown in Fig.~\ref{otocgraph} panel (f)---. In this case, positive- and negative-parity states are non-degenerate. However, there are states whose energy gaps with the adjacent states of opposite parity are very close. In such cases, some OTOC frequencies can again be very small, and thus the correlator may exhibit a long-period oscillation around zero, depending on the value of the matrix elements of the operators $\hat{V}$ and $\hat{W}$.

\medskip

\begin{figure*}
\begin{centering}
\includegraphics[scale=0.6]{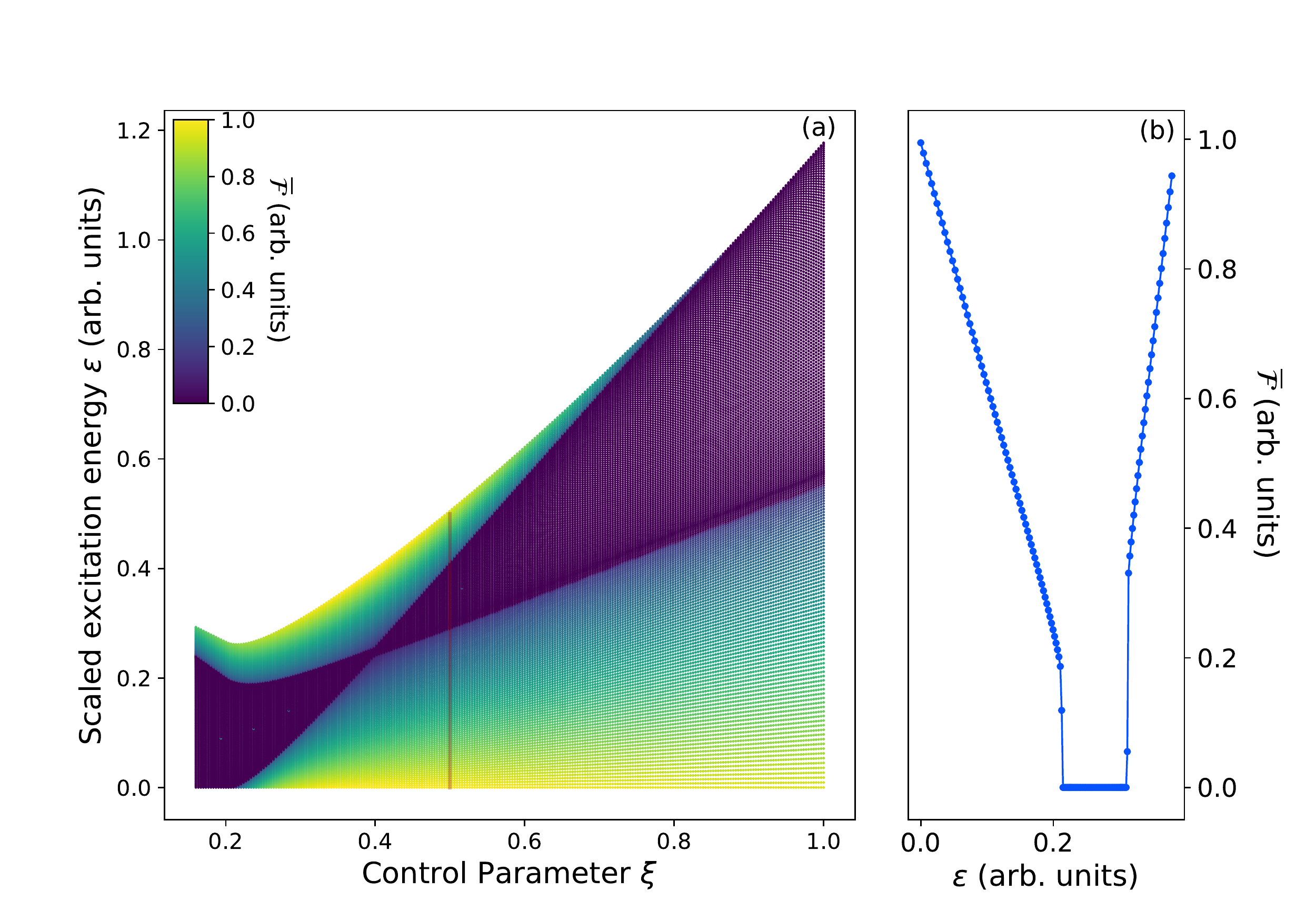}
\par\end{centering}
\caption{Panel a: Correlation energy diagram of the even parity states ALMG model as a function of the the control parameter $\xi$ with $\alpha = -0.6$. Each energy level is colored according to the steady-state value of the microcanonical OTOC,
  $\overline{{\cal F}}_{j}$, where $j$ takes
  the values $i = 0, 1, \ldots, N/2$. A vertical line marks the results for the $\xi = 0.5$ case whose $\overline{{\cal F}}_{j}$ are depicted in the right panel. Panel b: Steady-state value of the microcanonical OTOC,
  $\overline{{\cal F}_j}$, as a function of the re-scaled excitation energy
  $\overline{\varepsilon}$ for even parity states of a system with control parameters $\xi=0.5$ and $\alpha=-0.6$. Both panels: Calculations for a system with size $N = 400$.}
\label{otocall}
\end{figure*}

It is worth noticing that, in the anharmonic case, the two ESQPT
critical lines cross at $\xi=0.4$, as we can observe in
Fig. \ref{tangentgraph}. Since the  case depicted in
Fig. \ref{otocgraph} is for $\xi=0.5$, going up in energy from the
ground state the first ESQPT found is the anharmonic one (not present in the simple LMG model). We have checked, although
it is not shown here, that similar results are obtained if we consider
a value of $\xi$ such that the two critical lines have not crossed yet
($\xi-$value between $0.2$ and $0.4$). This is interesting because,
regardless the physical origin of the critical line, once the first
ESQPT critical line is crossed, the microcanonical OTOC starts oscillating around zero until the second ESQPT critical energy is
crossed. This has an immediate consequence on the steady-state value
of ${\cal F}_{i}(t)$ for $\hat{V}=\Spin_{-}/S$ and $\hat{W}=\Spin_{+}/S$, that can be taken as a reliable order parameter. This
is not the case for $\hat{W}=\hat{V}=\Spin_{x}/S$. It is true that
this choice of operators marks correctly the transition once the first
critical line is crossed, but it is not sensitive to the second ESQPT
critical line, irrespective of the physical origin.

\medskip

The steady state of the microcanonical OTOC ${\cal F}_{i}(t)$ is defined as 
\begin{eqnarray}
\overline{{\cal F}}_{i}&=&\lim_{T\to\infty}\frac{1}{T}\int_{0}^{T}{\cal F}_{i}(t)dt~.
\label{steady-state}
\end{eqnarray}
Results for this quantity can be found in Fig.~\ref{otocall}. In the left panel we depicted the correlation energy diagram for a system with size $N = 400$ as a function of the $\xi$ parameter for a fixed anharmonicity parameter value $\alpha=-0.6$. Each point is colored according to the corresponding $\overline{{\cal F}}_{i}$ value. From this figure it is clear how the stationary limit of the OTOC provides a convenient order parameter for the two ESQPTs in the ALMG model, with abrupt changes whenever the system gets through critical energies. The right panel of the same figure (Fig.~\ref{otocall}b) includes the stationary OTOC results for the eigenstates of an ALMG model with  system size $N=400$ and control parameters $\xi = 0.5$ and $\alpha=-0.6$  as a function of the re-scaled energy the rescaled excitation energy $\varepsilon_{n} = (E_n-E_{gs})/N$. This corresponds to the results marked with a vertical line in  Fig.~\ref{otocall}a. 
% \noindent where $T$ is the total evolution time. Besides, the
% re-scalede energy is defined as
% $\overline{\varepsilon}_{i}=2(E_i-E_0)/(E_{\mbox{max}}-E_0)$, $E_0$
% and $E_{\mbox{max}}$ are the ground state and the highest excited
% state eigenenergies, respectively.
\noindent The energy dependence of this quantity can be anticipated from the observation of the behavior of ${\cal F}_{i}(t)$ depicted in Fig. \ref{otocgraph}. Indeed, as pointed out previously, the main feature of ${\cal F}_{i}(t)$ is that it is an oscillatory function. However, the value around which it oscillates is different from zero only in the region below (above) the critical energy of the first (second)
ESQPT that is encountered when one goes up in energy in the
spectrum. Once the first ESQPT critical line is crossed, regardless of
its physical origin, the oscillations are around zero. This leads to
the conclusion that $\overline{{\cal F}}_{i}$ is different from zero in the region below the first critical line and it is equal to
zero in the region of the spectrum above that first ESQPT critical
line. However, it is worth noticing that for particular values of the control parameters there could  be nonzero instances of $\overline{{\cal F}}_{i}$. These nonzero values are isolated, akin to accidental degeneracies and are not associated to a critical energy, thus are easily distinguishable from the sudden change associated with the occurrence of an ESQPT.

% In Fig.~\ref{steadygraph}, the normalized steady-state value is
% depicted versus the re-scaled energies for a system size $N=300$. One
% can observe that $\overline{F}_{i}$ takes a value different from zero
% in the region just below the lowest critical line. For the case
% presented in Fig.~\ref{steadygraph}, the re-scaled critical energy for
% the lowest ESQPT is marked with $\varepsilon_{\mbox{c1}}=(1)$,
% meanwhile the re-scaled critical energy for the second ESQPT (higher)
% is marked with $\varepsilon_{\mbox{c2}}=(1)$. It is clear that for
% eigenstates in between the lowest and highest critical ESQPT energies,
% $\overline{F}_{i}$ is
% zero
%% A little bump is observed around $\overline{\varepsilon}=1.150(1)$ in the region between the two critical energies. This bump is actually an artifact, since the period of the oscillations for the envelope of $F_{n}(t)$ is bigger than the one found in the other regions. Therefore, when $\overline{F}_{n}$ is computed, there could exist small contributions that will vanish for larger integration time intervals.

In light of these results, it is clear that ESQPTs have a strong
impact on the OTOC dynamics, notwithstanding they can be mapped to a stationary point or the asymptotic behavior of the
PES in the semiclassical description of the system. Thus, the findings
given in Ref. \cite{CurroOTOC} are confirmed in this respect. However,
concerning the use of $\overline{{\cal F}}_{i}$ as an order parameter, we
have found that to be sensitive to both ESQPT lines the $\hat{W}$ and
$\hat{V}$ operators cannot be the same.

\section{Conclusions}

The ALMG model presents, in addition to the ground state QPT and its
known ESQPT, a second ESQPT. In this work, we have analyzed the impact of both ESQPT
critical lines on the dynamical evolution of the survival probability, the Loschmidt echo
and the OTOC. We have found that both ESQPTs, despite of having
different physical origins, lead to a dramatic change in the survival
probability evolution after a quantum quench. Particularly, it has been shown that the
survival probability gives information about system relaxation: for a
certain critical quench, related to the ESQPT energy, the system
behaves in a unique way that allows one to recognize the critical
lines separating regions in which the system is in a different
phase. In addition, it has been explained that due to the way we are
introducing the quantum quench it does not allow to reach with the
same procedure the two ESQPT lines that appear in the anharmonic
Lipkin model. Consequently, an alternative way for characterize one
ESQPT has been proposed. This method starts from the most excited
Hamiltonian eigenstate (instead of the ground state). Both
calculations are complementary and allow us to study the two ESQPT
lines. In both cases, the survival probability drops down to zero
(with small random fluctuations) when reaching an ESQPT line. This behavior was explained with the help of the LDOS.

\medskip

We show that other quantity whose evolution is greatly affected by the presence
of an ESQPT is the LE (see Figs.~\ref{fig:LE_t} and \ref{fig:LE_prod}). The time-dependence of the LE is quite different if the evolved state is close to the critical energy. Beyond the temporal evolution, the time-averaged LE has been proved to be a convenient ESQPT detector in the ALMG model too, since this quantity displays local maxima in the critical ESQPT energy values.

\medskip

An additional way of characterizing the dynamical evolution of the
system is the study of an OTOC. In this work, such a study has been
done using the microcanonical scheme and has revealed that the new
ESQPT, that is, the one generated by the anharmonicity term, also has
noticeable effects on the evolution of the OTOC. However, it is
difficult to determine sharply if the system has reached the critical
ESQPT energy, since there is a wide region close to the critical
energy where the system is affected by the corresponding
ESQPT. Finally, we have concluded that the normalized steady-state
value for ${\cal F}_{i}(t)$, $\overline{{\cal F}}_{i}$, can be used
as an order parameter to mark the two ESQPTs that occur in the ALMG,
despite the different nature of the two cases.

\begin{acknowledgments}
The authors thank José Enrique García Ramos, Miguel Carvajal Zaera, Ángel L. Corps and Armando Relaño for fruitful and inspiring discussions on the topic of this paper.
This work is part of the I+D+i projects PID2019-104002GB-C21,
PID2019-104002GB-C22, and PID2020-114687GB-I00 funded by
MCIN/AEI/10.13039/501100011033. This work has also been partially
supported and by the Consejer\'{\i}a de Conocimiento, Investigaci\'on
y Universidad, Junta de Andaluc\'{\i}a and European Regional
Development Fund (ERDF), refs.~UHU-1262561, PY2000764 (FPB), and US-1380840 and it is
also part of grant Groups FQM-160 and FQM-287 and the project PAIDI
2020 with reference P20\_01247, funded by the Consejería de Economía,
Conocimiento, Empresas y Universidad, Junta de Andalucía (Spain) and
“ERDF—A Way of Making Europe”, by the “European Union” or by the
“European Union NextGenerationEU/PRTR”. Computing resources supporting
this work were provided by the CEAFMC and Universidad de Huelva High
Performance Computer (HPC@UHU) located in the Campus Universitario el
Carmen and funded by FEDER/MINECO project UNHU-15CE-2848.
\end{acknowledgments}

\bibliography{refs}

\end{document}